\documentstyle[12pt]{article}
\begin{document}
\begin{titlepage}
\hspace{9cm} ULB--PMIF--92/01

\vspace{4.5cm}
\begin{centering}

{\huge Geometric Interpretation of the Quantum Master Equation in the
BRST--anti-BRST Formalism}\\
\vspace{1cm}
{\large Marc Henneaux$^*$\\
Facult\'e des Sciences, Universit\'e Libre de Bruxelles,\\
Campus Plaine C.P. 231, B-1050 Bruxelles, Belgium}\\
\end{centering}
\vspace{7cm}
{\footnotesize($^*$)Ma\^\i tre de Recherches au Fonds
National de la Recherche Scientifique. Also at Centro de Estudios
Cient\'\i ficos de Santiago, Chile.}
\end{titlepage}

\begin{abstract}
The geometric interpretation of the antibracket formalism given by Witten is
extended to cover the anti-BRST symmetry. This enables one to formulate the
quantum master equation for the BRST--anti-BRST formalism in terms of
integration theory over a supermanifold. A proof of the equivalence of the
standard antibracket formalism with the antibracket formalism for the
BRST--anti-BRST symmetry is also given.
\end{abstract}  \pagebreak
\section{Introduction}
The most powerful method for covariantly quantizing gauge theories is based
on the antibracket-antifield formalism \cite{deWit,Bat1,Bat2}\footnote{For
recent reviews developing the approach followed in this paper,
see\cite{MH1,book}.}. In that formalism, one associates with each field $
\Phi^A$   one antifield $\Phi^*_A$ of opposite Grassmann parity,

\begin{equation}
\epsilon(\Phi^*_A) = \epsilon_A + 1,\; \epsilon_A =
\epsilon(\Phi^A) .
\end{equation}
  One then solves the quantum master equation for the ``quantum action''\\
 $W(\Phi^A,\Phi^*_A)$,
\begin{equation}
i\Delta W -{1\over 2} (W,W) = 0 \Leftrightarrow \Delta(exp\,i\,W)=0,
\end{equation}
where we have set $\hbar=1$.  In (2), the operator $\Delta$ is defined by
\begin{equation}
\Delta = - (-)^{\epsilon_A}
\frac{\delta^R}{\delta\Phi^A}\frac{\delta^R}{\delta\Phi^*_A}
\end{equation}
and is nilpotent,
\begin{equation}
\Delta^2 = 0,
\end{equation}
while the antibracket is given by
\begin{equation}
(F,G) = \frac{\delta^R F}{\delta\Phi^A}\frac{\delta^L G}{\delta\Phi^*_A} -
\frac{\delta^R F}{\delta\Phi^*_A}\frac{\delta^L G}{\delta\Phi^A}.
\end{equation}
The generating functional of the Green functions is equal to
\begin{equation}
Z(j) = \int[D\Phi] exp\{i[W_\psi + j_A \Phi^A]\}
\end{equation}
with
\begin{equation}
W_\psi = W(\Phi,\Phi^* = \frac{\delta\psi}{\delta\Phi}).
\end{equation}
The vacuum functional $Z(0)$ does not depend on the choice of $\psi$ because
of the master equation.

As shown by Witten in \cite{Witten}, the antibracket formalism admits a
geometric interpretation if one identifies the antifields with the vectors
$\delta/\delta\Phi^A$ tangent to the coordinate lines in field space. The
function $W(\Phi,\Phi^*)$ can be seen as a multivector,
\begin{equation}
W(\Phi,\Phi^*) = W_0(\Phi) + W^A(\Phi)\Phi^*_A +\frac{1}{2} W^{AB}(\Phi)
\Phi^*_A \Phi^*_B + \cdots
\end{equation}
where $W_0(\Phi)$ is a scalar, $W^A(\Phi) \Phi^*_A$ a 1-vector, $\frac{1}{2}
W^{AB}(\Phi) \Phi^*_A \Phi^*_B$ a bivector...  The operator $\Delta$ becomes
the divergence of p-vectors, so that (2) expresses that $exp\,i\,W$ is
divergenceless. One can then interpret (6) in terms of integration theory
on a supermanifold  and the $\psi$-independence of $Z(0)$ is just
Stokes theorem in field space (see below),
\begin{equation}
\int\Delta F\, [D\Phi] = 0
\end {equation}
(assuming a fast decrease of $F$ at infinity in field space).

Recently, Batalin, Lavrov and Tyutin have extended the antibracket-antifield
formalism in order to include the anti-BRST symmetry
\cite{BLT1,BLT2,BLT3,L1}. To that end, they associate with each field not
just one antifield, but rather three antifields $\Phi^*_{Aa} (a=1,2)$ and
$\overline{\Phi}_A$ of parity \begin{equation}
\epsilon(\Phi^*_{Aa}) = \epsilon_A + 1,\; \epsilon(\overline{\Phi}_A) =
\epsilon_A. \end{equation}
They replace the quantum master equation (2) by
\begin{equation}
\frac{1}{2} (W,W)^a + V^a W = i \, \Delta^a W \;\;(a=1,2)
\end{equation}
with
\begin{equation}
(F,G)^a = \frac{\delta^R F}{\delta\Phi^A}\frac{\delta^L G}{\delta\Phi^*_{Aa}}
-\frac{\delta^R F}{\delta\Phi^*_{Aa}}\frac{\delta^L G}{\delta\Phi^A},
\end{equation}
\begin{equation}
V^a = \epsilon^{ab}\, \Phi^*_{Ab}\, \frac{\delta^L}{\delta\overline{\Phi}_A},
\;\;\epsilon^{ab} = - \epsilon^{ba}, \;\;\epsilon^{12} = 1,
\end{equation}
\begin{equation}
\Delta^a = - (-)^{\epsilon_A} \frac{\delta^R}{\delta\Phi^A}
\frac{\delta^R}{\delta\Phi^*_{Aa}}.
\end{equation}
In this extended formalism, the path integral having both BRST-invariance
and anti-BRST invariance is given by \cite{BLT1,BLT2}
\begin{equation}
Z(0) = \int [D\Phi] [\hat{U} exp\,i\,W]|_{\Phi^*_{Aa}=0,\overline{\Phi}_A=0}
\end{equation}
where the operator $\hat{U}$ is equal to
\begin{equation}
\hat{U} = exp[\,\frac{\delta^R F}{\delta\Phi^A}\,
\frac{\delta^L}{\delta\overline{\Phi}_A} + \frac{i}{2}\,\epsilon_{ab}\,
\frac{\delta^L}{\Phi^*_{Aa}}\, \frac{(\delta^R)^2 F}{\delta\Phi^A
\delta\Phi^B}\, \frac{\delta^L}{\delta\Phi^*_{Bb}}]
\end{equation}
and involves an arbitrary bosonic gauge fixing fermion $F(\Phi)$.

The purpose of this letter is to show that in spite of the increase in the
number of antifields, the formalism can still be given Witten's geometric
interpretation in terms of multivectors on the supermanifold of the fields.
The crucial new feature is that {\em the multivectors are now described in
an overcomplete basis}.  A byproduct of our analysis is a direct proof of
the equivalence of the extended BRST formalism based on (11) with the
original antibracket formalism based on the single quantum master equation
(2).

\section{Overcomplete Sets}
On the supermanifold M of the fields $\Phi^A$, let us introduce the
redundant basis of vectors
\begin{equation}
\Phi^*_{A1} = \frac{\delta}{\delta\Phi^A},\; \Phi^*_{A2} =
\frac{\delta}{\delta\Phi^A},
\end{equation}
that is, let us duplicate each basis vector $\delta/\delta\Phi^A$. The idea
of systematically duplicating ghost-like variables in the construction of the
anti-BRST symmetry has been introduced previously in \cite{Greg}. Let us also
introduce a vector bidegree that distinguishes between $\Phi^*_{A1}$ and
$\Phi^*_{A2}$ by setting
\begin{equation}
vect(\Phi^A) = (0,0);\; vect(\Phi^*_{A1}) = (1,0);\; vect(\Phi^*_{A2}) =
(0,1). \end{equation}
With the identification (17), every polynomial in $\Phi^A$, $\Phi^*_{A1}$ and
$\Phi^*_{A2}$ becomes a multivector on M. Conversely, every multivector on M
can be represented by at least one polynomial in $\Phi^A$, $\Phi^*_{A1}$ and
$\Phi^*_{A2}$.  However, that polynomial is not unique since both
$\Phi^*_{A1}$ and $\Phi^*_{A2}$ are identified with the same vector
$\delta/\delta\Phi^A$. Hence, a multivector has more than one expansion in
$\Phi^A$, $\Phi^*_{Aa}$. In order to identify the algebra of multivectors with
the algebra of polynomials in $\Phi^A$ and $\Phi^*_{Aa}$, it is necessary to
set $\Phi^*_{A1} - \Phi^*_{A2} =0$ in that latter algebra.

This task can be achieved by hand, but a more elegant procedure, used
repeatedly
in the BRST context \cite{book,FHST}, is to introduce a nilpotent operator
that does the job. The searched-for operator, denoted by $V$, is defined as
follows. First, one introduces extra variables $\overline{\Phi}_A$ of
degree $(1,1)$. Second, one sets
\begin{equation}
V \overline{\Phi}_A = \Phi^*_{A2}-\Phi^*_{A1},\; VJ\Phi^*_{Aa} = 0,\;
VJ\Phi^A = 0,
\end{equation}
i.e.,
\begin{equation}
V = V^1 + V^2,
\end{equation}
with $V^a$ given by (13), and one extends $V$ as a derivation. It is clear
that V is nilpotent, \begin{equation}
V^2 = 0,
\end{equation}
and that the cohomology of $V$ in the algebra of polynomials in $\Phi^*_{Aa}$
and $\overline{\Phi}_A$ with coefficients that are functions of $\Phi^A$, is
just the algebra of multivectors on $M$. Indeed, $\overline{\Phi}_A$ does
not contribute to the cohomology because it is not closed ($V
\overline{\Phi}_A \neq 0$), while $\Phi^*_{A2} - \Phi^*_{A1}$ is killed in
cohomology because it is exact.  Hence, only $\Phi^A$ and $\Phi^*_{A1} +
\Phi^*_{A2}$ (say) remain in the cohomology of $V$. One says that the
differential complex of polynomials in $\Phi^*_{Aa}$, $\overline{\Phi}_A$
with coefficients that are functions on $M$, equipped with the differential
$V$, is a {\em resolution} of the algebra of multivectors on $M$,
\begin{equation}
H^*(V) \simeq \{algebra\;\,of\;\,multivectors\;\,on\;\,M\}
\end{equation}
In that picture, the extra antifields $\overline{\Phi}_A$ of degree $(1,1)$
just appear as the variables that implement the identification of
$\Phi^*_{A2} -\Phi^*_{A1}$ with zero through the cohomology of $V$.

It is convenient to redefine the variables as
\begin{equation}
u^*_A = \frac{\Phi^*_{A2} + \Phi^*_{A1}}{2},\; v^*_A = \frac{\Phi^*_{A2}J-
\Phi^*_{A1}}{2}
\end{equation}
and to introduce a new degree such that $\Phi^A$ and $u^*_A$ are in degree
zero,
$v^*_A$ in degree one, and $\overline{\Phi}_A$ in degree two. In terms of
that degree, the cohomology of $V$ lies in degree zero and Equ. (22)
becomes \begin{equation}
H^0(V) = C^{\infty}(M) \otimes\,{\bf C}(u^*_A),\; H^k(V) = 0 \,(k\neq 0),
\end{equation}
since the algebra of multivectors on $M$ is isomorphic with the algebra \\
$C^{\infty}(M) \otimes\,{\bf C}(u^*_A)$.

\section{Divergence of Multivectors in the Overcomplete Representation}
We have seen that for multivectors, one can introduce the divergence operator
given by $$\Delta = -
(-)^{\epsilon_A}\frac{\delta^2}{\delta\Phi^A\,\delta(\delta/\delta\Phi^A)}.$$
Can one extend $\Delta$ to the algebra of polynomials in
$\Phi^*_{Aa}$ and $\overline{\Phi}_A$ with coefficients that are functions
on $M$ in such a way that its cohomology is unchanged? The answer is
affirmative, as it follows again from techniques familiar from BRST theory
\cite{FHST}. The required extension, which we denote by $\overline{\Delta}$
and still call the divergence operator, is given by
\begin{equation}
\overline{\Delta} = \overline{\Delta^1} + \overline{\Delta^2}
\end{equation}
with
\begin{equation}
\overline{\Delta^1} = \Delta^1 + i\, V^1, \; \overline{\Delta^2} = \Delta^2 +
 i\,V^2.
\end{equation}
That this is the correct $\overline{\Delta}$ can be seen as follows\footnote{
The inclusion of the factor $i$ in (26) is purely conventional
and can be absorbed in a
redefinition of the variables. Our conventions follow those of
\cite{BLT1,BLT2}}.
\begin{enumerate} \item It is easily verified that $\overline{\Delta}$ is
nilpotent,
\protect\begin{equation}
\overline{\Delta}^2 = 0.
\protect\end{equation}
\item In terms of the degree introduced in Equ.(24), $\overline{\Delta}$
splits as
\protect\begin{equation}
\overline{\Delta} = \tilde{\Delta} + i\, V
\protect\end{equation}
where $V$ is in degree $-1$ and $\tilde{\Delta} = \Delta^1 + \Delta^2$ is in
degree
$0$. By standard spectral sequence arguments, the cohomology of $\overline
{\Delta}$ is
given by the cohomology of the operator induced by $\overline{\Delta}$ in the
cohomology $H^*(V)$ of $V$, that is, it is the cohomology of the divergence
operator
in the algebra of multivectors on $M$, as required.
\end{enumerate}
The two equations in (11) imply\footnote{The single
equation (29) implies in turn the two equations in (11) if one requires $W$ to
be of appropriate ghost bidegree, see \cite{Greg2} for more information. This
point
will not be needed in the
sequel.}
\begin{equation}\overline{\Delta}\, exp\, i\, W = 0
\end{equation}
Thus, the quantum master equation of the extended BRST-anti-BRST formalism
expresses that the multivector $exp\,i\,W$ is divergencefree, just as in the
ordinary case considered in \cite{Witten}.

\section{Integration Theory}
The geometric point of view enables one to get a better understanding of the
formalism. In particular, it sheds a new light on why the path integral (15)
does
not depend on the choice of the gauge fixing function and coincides with the
path integral $Z(0)$ (Equ.(6)) of the non-extended formalism.

The natural definition of the integral of a multivector $A$ over $M$,
\begin{equation}
A = A_0 + A_1 + A_2 + \dots
\end{equation}
(where $A_0$ is the scalar part of $A$, $A_1$ its $1$-vector part \dots)
consists in setting
\begin{equation}
\int A\, [D\Phi] \equiv  \int A_0\,[D\Phi],
\end{equation}
where $\int A_0\,[D\Phi]$ is the standard integral of functions
\footnote{We assume that a measure (superdensity of weight one) has been given
and is equal to $1$ in the given $\Phi^A$-coordinate system on $M$.}. Thus,
the integral of a $p$-vector over $M$ is zero if $p \neq 0$. One can rewrite
(31) as
\begin{equation}
\int A\, [D\Phi] \equiv \int A(\Phi,\Phi^* = 0,\overline{\Phi} = 0).
\end{equation}
A crucial property of the integral (31) is given by Stokes theorem,
\begin{equation}
\int divergence(B)\,[D\Phi] = 0,
\end{equation}
where B is an arbitrary multivector
decreasing fast enough at infinity in field space. Equ. (33) plays a
fundamental role in
quantum field theory and contains the Schwinger-Dyson equations  (see e.g.
\cite{book}). In terms of the redundant description of multivectors, (33)
becomes
\begin{equation}
\int \overline{\Delta}\,B\,[D\Phi] = 0.
\end{equation}
Because of Stokes theorem, the integral is defined in the
cohomology algebra $H^*(\overline{\Delta})$ of the divergence operator
$\overline{\Delta}$.

The simple definition (31) is too restrictive
and must be generalized for gauge theories. Indeed, one finds in that case
that the integral of $exp\,i\,W$ is ill-defined: the integral over the gauge
modes and the ghosts yields $0\:\delta(0)$. The way to make sense out of the
integral of $exp\,i\,W$ is to regularize it by means of a total divergence.
More precisely, one replaces $exp\,i\,W$ by  \begin{equation}
\hat{U}_K\;exp\,i\,W,
\end{equation}
\begin{equation}
exp\,i\,W \longrightarrow \hat{U}_K\;exp\,i\,W,
\end{equation}
where $\hat{U}_K$ is the operator
\begin{equation}
\hat{U}_K = exp\,[\hat{K},\,\overline{\Delta}].
\end{equation}
One has
\begin{equation}
exp\,[\hat{K},\,\overline{\Delta}] = 1 + [\hat{J},\,\overline{\Delta}]
\end{equation}
from which it follows
\begin{equation}
\hat{U}_K\;exp\,i\,W = exp\,i\,W + \overline{\Delta}\:(\hat{J}\:exp\,i\,W)
\end{equation}
for some operator $\hat{J}$ since $\overline{\Delta}$ is nilpotent and
$exp\,i\,W$ is $\overline{\Delta}$-closed.

Because the operator $\hat{U}_K$ differs
from the identity by a $\overline{\Delta}$-exact operator, one defines
more generally the integral of a $\overline{\Delta}$-closed multivector A by
\begin{equation} \int [D\Phi]\,A \equiv \int[D\Phi]\,(\hat{U}_K\,A)_0
\end{equation} where the operator $\hat{K}$ determines the gauge fixing
procedure and is chosen in such a way that the right-hand side of (40) is
well-defined\footnote{If the right-hand side of (31) is already well-defined,
one can take $\hat{K} = 0$. For $\int [D\Phi]\,exp\,i\,W$ in a gauge theory,
one must take $\hat{K}$ different from zero.}. This definition of the
integral possesses the following good properties: \begin{enumerate}
\item It reduces to the usual definition (31) when this latter is applicable.
\item Because of Stokes theorem, (40) is formally independent on the choice of
$\hat{K}$ (``Fradkin-Vilkovisky theorem''). Indeed, one has
\protect\begin{equation}
\hat{U}_{K'}\,A = \hat{U}_K\,(\,1 + [\tilde{J}_{K,K'},\,\overline{\Delta}])
\,A =
\hat{U}_K\,A + \overline{\Delta}C
\protect\end{equation}
for some $C$, since $\overline{\Delta} A = 0$.
\item The integral is actually a function of the cohomological classes of
$\overline{\Delta}$, since $\hat{U}_K$ commutes with $\overline{\Delta}$.

\end{enumerate}
For these reasons, the definition (40) is quite reasonable.

Now, the integral (15) considered inJ\cite{BLT1,BLT2} {\em is precisely of the
form} (40). Indeed, one can write (16) as
\begin{equation}
exp\,[\hat{K},\,\overline{\Delta}]
\end{equation}
with
\begin{equation}
i\,\hat{K} = \frac{\delta\,F}{\delta\Phi^B}\,\frac{\delta}{\delta\Phi^*_{B1}}
- \frac{\delta\,F}{\delta\Phi^B}\,\frac{\delta}{\delta\Phi^*_{B2}}.
\end{equation}
The path integral (15) is thus quite natural from the point of view of the
integration theory on $M$. It fits within the above construction and can be
viewed
as a regularization of the naive definition (31) by means of the addition of a
particular $\overline{\Delta}$-exact term.

 The choice of $\hat{K}$ in (43) yields
a path integral that is both BRST and anti-BRST invariant \cite{BLT1,BLT2}.
However, one may take a gauge fixing $\hat{K}$ that has a different structure
without changing the integral. For instance, one may take a $\hat{K}$ that
provides a direct link with the original antifield formalism. This is done
through \begin{equation}
\hat{K} = - \psi(\Phi^A)
\end{equation}
where $\psi$ is the operator of multiplication by $\psi$. One has
\begin{equation}
[\hat{K},\,\overline{\Delta}] =
\frac{\delta\psi}{\delta\Phi^A}\,\frac{\delta}{\delta u^*_A}
\end{equation}
and thus
\begin{equation}
\hat{U}_K\,A = A(\Phi,\,u^* + \frac{\delta\psi}{\delta\Phi}, \,v^*,
\,\overline{\Phi}).
\end{equation}
This yields the path integral (6) if one identifies $u^*_A$ with $\Phi^*_A$.
Indeed, the function $W(\Phi,\,u^* = \Phi^*,\,v^* = 0, \,\overline{\Phi} =
0)$ is a solution of the quantum master equation (2). This is a direct
consequence of (11).  Furthermore,\\ $W(\Phi,\,u^* = \Phi^*,\,v^* = 0,\,
\overline{\Phi} = 0)$ can easily be seen to be a (non minimal) {\em proper
solution}
of (2); the ghost spectrum of \cite{BLT1,BLT2} merely corresponds to a
duplication
of the gauge symmetries \cite{Greg2}. Hence, there is complete equivalence
with the
standard formalism, because the path integral does not depend on the choice
of the non minimal sector.  For another equivalence proof, see \cite{BLT3}.

\section{Conclusion}
We have shown in this letter that the antifield formalism for the combined
BRST-anti-BRST symmetry developed in \cite{BLT1,BLT2,BLT3,L1} can be given a
geometric interpretation in terms of multivectors on the supermanifold of the
fields. This extends the work of
\cite{Witten} to the anti-BRST context.  The crucial point was to introduce a
redundant description of the multivectors and to introduce further variables
($\overline{\Phi}^A$) that kill the redundancy in cohomology. We have also
given a
direct geometrical proof of equivalence with the standard antifield formalism
in
terms of integration theory.

\section{Acknowledgements}
This work has been supported in part by a research contract with the
Commission of the European Communities.

\pagebreak

\end{document}